\documentstyle[epsfig,12pt,dina4h]{article}
\begin{document}                                                         
\newcommand{\kl}       {\mbox{$ K^0_L$}}
\newcommand{\costh}       {\mbox{$ cos \theta_h$}}
\newcommand{\k}        {\mbox{$ K^0$}}
\newcommand{\gsim}      {\mbox{\raisebox{-0.4ex}{$\;\stackrel{>}{\scriptstyle \sim}\;$}}}
\newcommand{\kb}       {\mbox{$ \overline{K^0}$}}
\newcommand{\lam}      {\mbox{$\Lambda$}}
\newcommand{\perc}      {~\mbox{\hspace*{-0.3em}\%}}
\newcommand{\lb}       {\mbox{$\overline{\Lambda}$}}
\newcommand{\GeV}      {\mbox{\rm GeV}}
\newcommand{\GeVc}     {\mbox{\rm GeV/c}}
\newcommand{\mpipi}    {\mbox{${M_{\pi\pi}}$}}
\newcommand{\mppi}     {\mbox{${M_{p\pi}}$}}
\newcommand{\mee}      {\mbox{${M_{ee}}$}}
\newcommand{\delz}     {\mbox{$\bigtriangleup Z$}}
\newcommand{\angxy}    {\mbox{$\alpha_{{\small XY}}$}}
\newcommand{\ctau}     {\mbox{$\rm c\tau$}}
\newcommand{\ra}       {\mbox{$\rightarrow $}}
\newcommand{\ptr}      {\mbox{$p_{\rm T}$}}
\newcommand{\pms}      {\mbox{$\pm$}}
\newcommand{\dspm}       {\mbox{$ D^{*\pm}$}}
\newcommand{\ds}       {\mbox{$ D^{*}$}}
\newcommand{\etamx}    {\mbox{$\eta_{\rm max}$}}
\newcommand{\pspu}    {\mbox{$P_s/P_u$}}
\newcommand{\psupudpspd}   {\mbox{$(P_{su}/P_{ud})/(P_s/P_d)$}}
\newcommand{\pqqpq}    {\mbox{$P_{qq}\P_q$}}
\begin{titlepage}
\title {
{\bf
\vspace*{3.cm}  {\boldmath $D^\ast$}  Production in Deep Inelastic Scattering \\
                at HERA } \\
\vspace*{2.cm}
 }

\author{ ZEUS Collaboration }
\date{}
\maketitle
\vspace{3 cm}

\begin{abstract}
 This paper presents measurements of \dspm\ production 
in deep inelastic scattering from collisions 
between 27.5\,GeV positrons and 820\,GeV protons.
The data have been taken with the ZEUS detector at HERA. The decay channel 
$D^{\ast +}\rightarrow (D^0 \rightarrow K^- \pi^+) \pi^+ $ (+\,c.c.) has
been used in the study. The $e^+p$ cross section for inclusive \dspm\ production
with $5<Q^2<100\,$GeV$^2$ and $y<0.7$ 
is 5.3\,\pms\,1.0\,\pms\,0.8\,nb
in the kinematic region {$1.3<\ptr(\dspm)<9.0$\,GeV and
$\left|\,\eta(\dspm)\,\right|<1.5$}.
Differential cross sections as functions   of \ptr(\dspm), $\eta$(\dspm),
$W$ and $Q^2$ are compared 
with next-to-leading order QCD calculations
based on the photon-gluon fusion production mechanism. 
After an extrapolation of the cross section to the full
kinematic region in \ptr(\dspm) and $\eta$(\dspm), the charm
contribution $F_2^{c\bar{c}}(x,Q^2)$ to the proton
structure function is determined for Bjorken $x$ between 2\,$\cdot$\,10$^{-4}$
and 5\,$\cdot$\,10$^{-3}$.
\end{abstract}


\vspace{-19.5 cm}
\setcounter{page}{0}
\thispagestyle{empty}
\end{titlepage}

\newpage

\topmargin-1.cm                                                                 
                   
\evensidemargin-0.3cm                                                           
                   
\oddsidemargin-0.3cm                                                            
                   
\textwidth 16.cm                                                                
                   
\textheight 680pt                                                               
                   
\parindent0.cm                                                                  
                   
\parskip0.3cm plus0.05cm minus0.05cm                                            
                   
\def\3{\ss}                                                                                        
\newcommand{\address}{ }                                                                           
\pagenumbering{Roman}          

                                                   %
\begin{center}                                                                                     
{                      \Large  The ZEUS Collaboration              }                               
\end{center}                                                                                       
  J.~Breitweg,                                                                                     
  M.~Derrick,                                                                                      
  D.~Krakauer,                                                                                     
  S.~Magill,                                                                                       
  D.~Mikunas,                                                                                      
  B.~Musgrave,                                                                                     
  J.~Repond,                                                                                       
  R.~Stanek,                                                                                       
  R.L.~Talaga,                                                                                     
  R.~Yoshida,                                                                                      
  H.~Zhang  \\                                                                                     
 {\it Argonne National Laboratory, Argonne, IL, USA}~$^{p}$
% ...
\newpage                                                                                           
% ...
$^{\    2}$ now at Univ. of Salerno and INFN Napoli, Italy \\                                      
$^{\    3}$ now at Univ. of Crete, Greece \\                                                       
$^{\    4}$ supported by Worldlab, Lausanne, Switzerland \\                                        
% ...
$^{  22}$ now at ATLAS Collaboration, Univ. of Munich \\                                           
% ...
                                                           %
                                                           %
\newpage   
                                                           %
                                                           %
\begin{tabular}[h]{rp{14cm}}                                                                       
$^{a}$ &  supported by the Natural Sciences and Engineering Research                               
          Council of Canada (NSERC)  \\                                                            
$^{b}$ &  supported by the FCAR of Qu\'ebec, Canada  \\                                            
$^{c}$ &  supported by the German Federal Ministry for Education and                               
          Science, Research and Technology (BMBF), under contract                                  
          numbers 057BN19P, 057FR19P, 057HH19P, 057HH29P, 057SI75I \\                              
$^{d}$ &  supported by the MINERVA Gesellschaft f\"ur Forschung GmbH,                              
          the German Israeli Foundation, and the U.S.-Israel Binational                            
          Science Foundation \\                                                                    
$^{e}$ &  supported by the German Israeli Foundation, and                                          
          by the Israel Science Foundation                                                         
  \\                                                                                               
$^{f}$ &  supported by the Italian National Institute for Nuclear Physics                          
          (INFN) \\                                                                                
$^{g}$ &  supported by the Japanese Ministry of Education, Science and                             
          Culture (the Monbusho) and its grants for Scientific Research \\                         
$^{h}$ &  supported by the Korean Ministry of Education and Korea Science                          
          and Engineering Foundation  \\                                                           
$^{i}$ &  supported by the Netherlands Foundation for Research on                                  
          Matter (FOM) \\                                                                          
$^{j}$ &  supported by the Polish State Committee for Scientific                                   
          Research, grant No.~115/E-343/SPUB/P03/120/96  \\                                        
$^{k}$ &  supported by the Polish State Committee for Scientific                                   
          Research (grant No. 2 P03B 083 08) and Foundation for                                    
          Polish-German Collaboration  \\                                                          
$^{l}$ &  partially supported by the German Federal Ministry for                                   
          Education and Science, Research and Technology (BMBF)  \\                                
$^{m}$ &  supported by the German Federal Ministry for Education and                               
          Science, Research and Technology (BMBF), and the Fund of                                 
          Fundamental Research of Russian Ministry of Science and                                  
          Education and by INTAS-Grant No. 93-63 \\                                                
$^{n}$ &  supported by the Spanish Ministry of Education                                           
          and Science through funds provided by CICYT \\                                           
$^{o}$ &  supported by the Particle Physics and                                                    
          Astronomy Research Council \\                                                            
$^{p}$ &  supported by the US Department of Energy \\                                              
$^{q}$ &  supported by the US National Science Foundation \\                                       
\end{tabular}                                                                                      
\clearpage
                                                           %

\pagenumbering{arabic}          

\setcounter{page}{1}

\section{\bf Introduction}
\label{s:Intro}

\begin{figure}
\epsfig{file=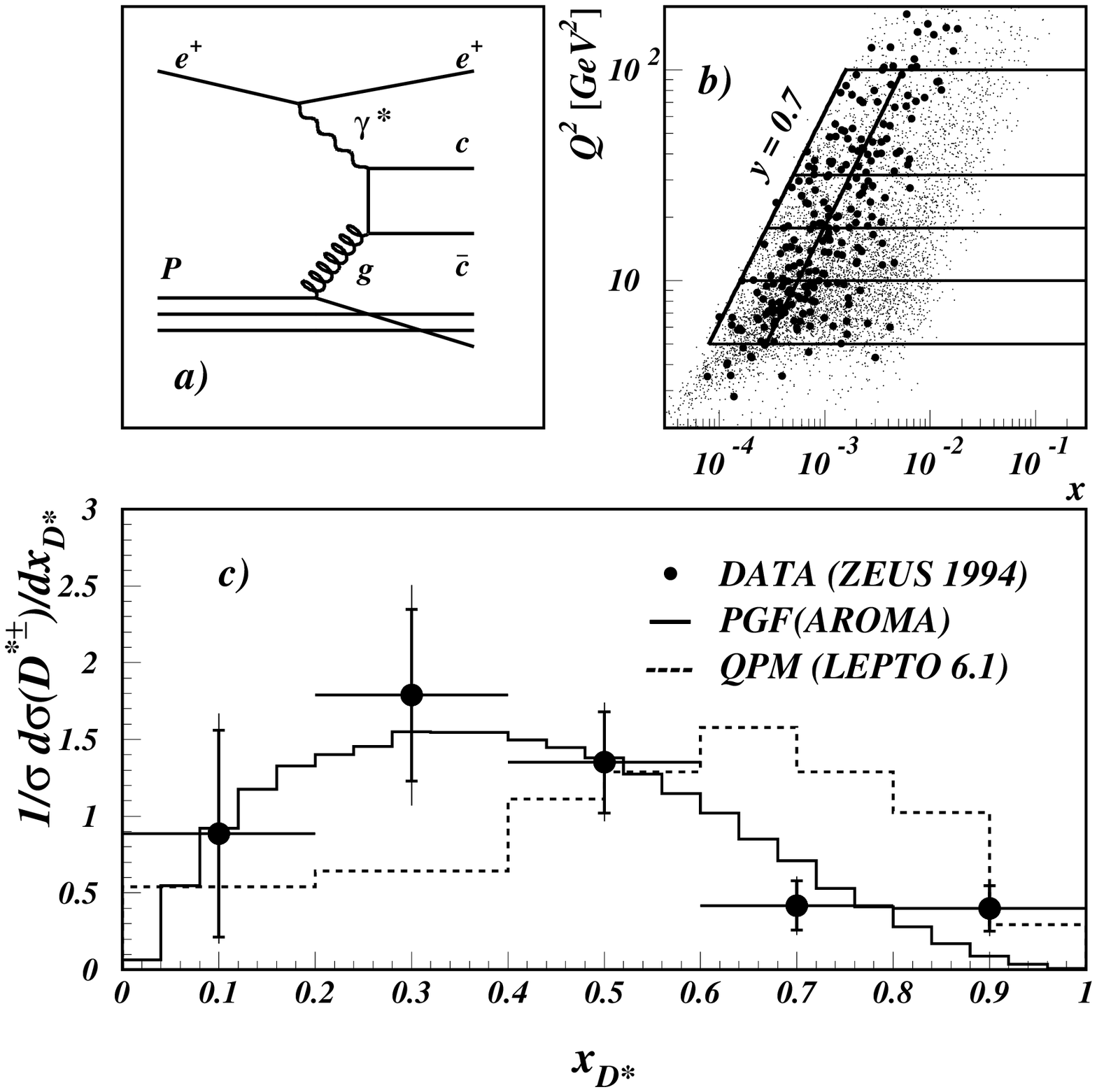,%
bbllx=2cm,bblly=150pt,bburx=525pt,bbury=600pt,height=16cm}
{\small
\caption{\label{fig:fig_bgf+} \em (a) LO diagram for 
photon-gluon fusion (PGF). (b) The ($x,Q^2$) plane with the
$Q^2$, $y$ region and bins 
chosen for the $F_2^{c\bar{c}}$ analysis. Large
dots correspond to the $D^\ast$ candidates; small dots correspond to
a subsample of DIS events. The second constant $y$ line corresponds to $y=0.2$.
(c) Normalized $e^+p \rightarrow e^+\dspm X$ cross section
for $5<Q^2<100$\,GeV$^2$, 
$y<0.7$ in the restricted kinematic region 
$1.3<\ptr(\ds)<9.0$\,GeV 
and $|\eta(\ds)|<1.5$
as a function of $x_{D^\ast}$.
The inner error bars show the statistical errors and the outer ones
the statistical and systematic errors added in quadrature.
The horizontal bars represent the bin widths. 
The prediction for PGF as calculated with 
AROMA (solid histogram) and the charm sea contribution
as calculated with LEPTO\,6.1 selecting QPM events (dashed histogram) 
are also shown (see text). }
}
\end{figure}

In neutral current deep inelastic scattering (DIS), $ep \rightarrow eX$,
charmed quarks are expected to be produced predominantly 
via the photon-gluon fusion (PGF) process which couples
the virtual photon to a gluon of the proton.
The leading order (LO) diagram is shown 
in Fig.\,\ref{fig:fig_bgf+}a. 

Recently, analytic calculations of the DIS charm cross section from photon-gluon
coupling have become available\,\cite{HARRIS}, which
relate the DIS charm cross section to the gluon distribution in the proton
using next-to-leading order (NLO) QCD.

Measurements of deep inelastic neutral current scattering at
HERA have demonstrated a rapid rise of the proton structure
function $F_2$ as Bjorken-$x$ decreases below $10^{-2}$\,\cite{Z93F2}. 
A QCD analysis of these data has connected this rise to an increase
of the gluon momentum density in the proton with a dependence
$x^{-\lambda}$ with $\lambda\,=\,0.35\,^{+0.04}_{-0.10}$ at 
$Q^2\,=\,7$\,GeV$^2$\,\cite{GZ,GH}.
At HERA, the DIS charm cross section is sensitive to the gluon distribution
of the proton at low fractional momentum  ($x_g \sim 10^{-3}$)
of the proton.
Comparison of the DIS charm cross section with NLO QCD calculations
allows an independent check of the increase of the gluon momentum density,
testing the consistency of the QCD calculations.

The other processes, apart from PGF, that contribute to open charm
production in DIS are: 
diffractive heavy flavour production \cite{DIFF},
scattering of the virtual photon off the charm sea quark \cite{IC}, 
charmed hadron production from $b\bar{b}$\,\cite{AMIO}
and production  of 
$c\bar{c}$ in fragmentation\,\cite{SPM}. These processes, however,
are expected to have much smaller cross sections than PGF.
The possible contribution from intrinsic
charm \cite{ICH,IC} is outside the acceptance of the main detector.

ZEUS \cite{RORI,DSPHP2}\,and H1 \cite{H1} have 
reported on $D^\ast$ production\footnote{In 
this paper charge conjugate modes
are always implied. $D^\ast$ always refers
to both $D^{\ast+}$ and the charge
conjugate mode (c.c.) $D^{\ast-}$.}
by quasi-real photons at HERA (i.e. in the photoproduction regime). 
In \cite{RORI} first $D^\ast$ signals in the DIS regime were also shown.

In this paper a detailed study of the $D^\ast$ 
produced in
DIS events is presented. $D^\ast$ are investigated in the decay channel 
\begin{equation}
D^{\ast +}\rightarrow D^0 \pi^+_s \rightarrow K^- \pi^+ \pi^+_s  (+\,c.c.)
\end{equation}
using a procedure first proposed in \cite{DDC}, where $\pi^+_s$ 
stands for the `soft pion'.

The $e^+p$ cross section for inclusive \dspm\, production
and differential cross sections as functions of \ptr($D^\ast$),
 $\eta$($D^\ast$), $W$ and $Q^2$ are  presented. 
The measurements are compared with LO and NLO
QCD analytic calculations \cite{HARRIS}
based on the photon-gluon fusion production mechanism.
By extrapolation, 
the charm contribution to the proton
structure function $F_2$, $F_2^{c\bar{c}}(x,Q^2)$, is estimated and compared 
with the NLO QCD analytic calculations.
A similar analysis has been recently presented  by the H1 
Collaboration \cite{MEMEO}.

\section{\bf Experimental setup}

The data were collected at the positron-proton collider HERA 
using the ZEUS detector during the 1994 running period. HERA collided 
27.5\,GeV positrons with 820\,GeV protons yielding a center-of-mass
energy of 300\,GeV. 
153 bunches were filled for each beam, and, in addition, 15 positron 
and 17 proton bunches were left unpaired for background studies.
The r.m.s. of the vertex position distribution in the $Z$ 
direction\footnote{The 
ZEUS coordinate system is defined as right-handed 
with the $Z$ axis pointing in the proton beam direction and the $X$ axis 
horizontal pointing towards the center of HERA. The polar angle $\theta$ 
is defined with respect to the positive $Z$ direction.} is 12\,cm.
The data used in this analysis come from
an integrated luminosity of 2.95\,pb$^{-1}$.

ZEUS is a multipurpose detector which
has been described in detail elsewhere \cite{ZEUS1}. 
The key component for this analysis is the central tracking detector
(CTD) which operates in a magnetic field of 1.43 T provided by a
thin superconducting solenoid. The CTD is a cylindrical drift chamber
consisting of 72 layers
organized into 9 ``superlayers'' covering the polar 
 angular
region $15^\circ < \theta < 164^\circ$\cite{CTD}. 
Five of the superlayers
have wires parallel (axial) to the beam
axis and four have wires inclined at a small angle to give a stereo view. 
The spatial resolution in the drift direction is 190\,$\mu$m.
The interaction vertex is measured with a resolution of 
0.4\,cm in the $Z$ direction and 0.1\,cm in the $XY$ plane. 
The momentum resolution for tracks traversing 
all 9 superlayers is 
$\sigma(p_{\rm T}) / p_{\rm T} = 
 0.005\, p_{\rm T} \bigoplus 0.016 $ ($p_{\rm T}$ in GeV).

The solenoid is surrounded by a high resolution uranium-scintillator 
calorimeter (CAL) described elsewhere \cite{CALRes}.
The position of positrons scattered close to the positron beam 
direction is determined by a scintillator strip detector (SRTD)
\cite{SFTVTX}. 
The luminosity is measured via the Bethe-Heitler process, $ e p \rightarrow
e p \gamma$, where the photon is tagged 
using a lead-scintillator calorimeter\,\cite{LUMIhard}
located at $Z\,=\,-\,107$\,m in the HERA tunnel. 

\section{\bf HERA kinematics}

The kinematics of deep inelastic scattering processes at HERA,
$e^+(k) + p(P) \rightarrow e^+(k^\prime) + X$,  
where $X$  is the hadronic final state, can be described by the Lorentz 
invariant variables $Q^2$, $x$ and $y$.  
Here $-Q^2$ is the square of the four-momentum
transfer at the lepton vertex,
$x$ is the Bjorken variable and
$y$ is the fractional energy transfer between the positron and the proton in 
the proton rest frame.
In the absence of QED radiation,
${Q^2\,=\,-q^2\,=\,-(k-k^\prime)^2\,,}$ and $x\,=\,{Q^2\,\over\, 2P\, \cdot\, q}$,
where $k$ and $P$ are the four-momenta of the incoming particles and
$k^\prime$ is the four-momentum of the scattered positron.
The variables  are related 
by $ Q^2 = s x y $, where $s$\ is the squared invariant 
mass of the $ep$ \ system.  
Since the ZEUS detector is nearly hermetic, for neutral current DIS
$Q^2$, $x$ and $y$ can be calculated from the kinematic variables of the
scattered positron, from the hadronic final state variables, or from a 
combination of both. In this paper we use the 
double angle method (DA) \cite{Q2DA} 
to calculate the $Q^2$, $x$, $y$ variables. 
The center-of-mass energy of the
virtual photon-proton system ($\gamma^* p$), $W$, is determined using
$W^2_{DA} = m_p^2 + Q^2_{DA} ({1 \over x_{DA}} - 1), $
$m_p$ being the proton mass.
The variable $y$, determined  using the energy $E^{\prime}_e$ and
angle $\theta^{\prime}_e$ of the scattered positron,
$y_e  =  1 -  {E^{\prime}_e \over 2 E_e} \ (1-\cos \theta^{\prime}_e)$,
is used for background suppression.
The use of the                                       
positron information alone to calculate $Q^2$, $x$ and $y$ 
is used in the study of the systematic uncertainties.

Using the calorimeter information the quantity $\delta= \sum_i (E_i-p_{Z_i})$ 
is measured, where $E_i$ is the energy and $p_{Z_i}$ the longitudinal 
momentum assigned to the calorimeter cell $i$.
For perfect detector resolution and acceptance, 
$\delta\,=\,55$\,GeV for DIS events while for photoproduction events
and events with  hard initial state radiation,
where the scattered positron or the radiated photon 
escapes down the beam pipe,
$\delta$ peaks at lower values.

\section{\bf DIS event selection}
\subsection{\bf Trigger selection}

The trigger selection is almost identical to that used for the 
measurement of the structure function $F_2$~\cite{F294}. 
Events are filtered online by a three level trigger 
system~\cite{b:CALFLT}. 
At the first level DIS events are selected by requiring a minimum
energy deposition in the electromagnetic section of the CAL.
The threshold depends on the position in the CAL and varies between
3.4 and 4.8 GeV.
At the second level trigger (SLT), 
beam-gas background is further reduced using the measured 
times of energy deposits and the summed energies from the calorimeter. 
The events are accepted if $\delta$ calculated at the SLT level
using the nominal vertex position satisfies
$  \delta_{SLT} > 24 \:\:{\rm GeV} - 2E_{\gamma},$
where $E_{\gamma}$
is the energy deposit measured in the luminosity photon calorimeter.

The full event information 
is available at the third level trigger (TLT).
Tighter timing cuts as 
well as algorithms to remove beam halo muons and cosmic muons are
applied.
The quantity $\delta_{TLT}$ is determined in the same manner as for
$\delta_{SLT}$. The events are required to have
$\delta_{TLT} > 25 \:\:{\rm GeV} - 2E_{\gamma}$.
Finally, events are accepted as DIS candidates if
a scattered positron candidate of energy greater than 4~GeV is found.
For events with the scattered positron detected in the calorimeter, 
the trigger acceptance is essentially independent of the DIS hadronic 
final state. It is greater than 90\% for $Q^2\,\approx$\,5\,GeV$^2$
(lower limit for the DIS sample in this analysis)
and increases to 99\,\% for $Q^2\,>$\,10\,GeV$^2$\, 
as determined from MC simulation. 

\subsection{\bf Offline event selection}

The selection of DIS events is similar to
that described in our earlier publication \cite{F294}. 
The characteristic signature of a DIS event is the scattered positron
detected in the uranium scintillator calorimeter. 
The positron identification algorithm is based on a neural network using
information from the CAL and is described elsewhere \cite{MORRA}.
The efficiency for finding the scattered positron is sensitive
to details of the shower evolution, in particular to energy loss
in the material between the interaction point and the calorimeter.
The efficiency of the identification algorithm when the scattered
positron has an energy of 8\,GeV is $50 \%$, rising to 99\,$\%$
for energies above 15\,GeV. For the present data, 89\,$\%$ of the events have
an $e^+$ energy greater than 15\,GeV.
The impact position of the positrons is determined by either the
position reconstructed by the CAL
or by the SRTD if inside its fiducial volume.
The resolution of the impact position is about 1\,cm for the CAL
and 0.3\,cm for the SRTD.
The positron traverses varying amounts of material in the detector 
before entering the CAL, which causes a variable energy loss. 
This energy loss is corrected on an event by event basis, 
as explained in\,\cite{SFTVTX}.
The scale uncertainty on the energy of the 
scattered positron after these corrections
is 2\perc\ at 10 GeV linearly decreasing to 1\perc\ at 27.5 GeV.

The following criteria are used to select DIS events:
the presence of a scattered positron candidate with 
a corrected energy $E^{\prime}_e >8$\,GeV, and
$ \delta >$ 35\,GeV to remove
photoproduction events and to 
suppress events with hard initial state radiation.
After applying these criteria, $4.3\cdot 10^{5}$ events are retained. 
For this analysis the region
5$<Q^2<$100\,GeV$^2$ and $y_e<0.7$ is selected which
contains $3.7\cdot 10^{5}$  events. 
A subsample of DIS events in the ($x,Q^2$) plane is plotted in
Fig.\,\ref{fig:fig_bgf+}b. The lines of
constant $y$ and $Q^2$ delimiting 
the region chosen for this study are shown.
The photoproduction contamination is less
than 2\% and the beam gas background is negligible.

\subsection{\bf Event selection efficiency}

The efficiencies of the event selection and the 
$D^\ast$ reconstruction are determined
using a GEANT \cite{GEANT} based Monte Carlo\,(MC) simulation 
program which incorporates the knowledge 
of the detector and the trigger. 
For this analysis two different types of 
event generators were used:
neutral current (NC) DIS generators, discussed below,  for 
event selection efficiency calculations and heavy flavour generators, discussed
in section 5.2, for $D^\ast$ reconstruction efficiency estimates.

Neutral current DIS events with $Q^2>4$\,GeV$^2$ were 
generated using the
HERACLES 4.4 program \cite{SPIES} which incorporates first order
electroweak corrections. 
It was interfaced using DJANGO\,6.1-2 \cite{DJANGO}
to either LEPTO 6.1-3 \cite{INGEL} or ARIADNE 4.03 \cite{ARI} 
for the simulation of QCD cascades. 
The calculation of the zeroth and first order
matrix elements plus the parton shower option 
(MEPS) was used in LEPTO. The latter 
includes coherence effects in the final 
state cascade via angular ordering of successive parton emissions.
In ARIADNE, the colour-dipole model including the photon-gluon fusion
process (CDMBGF) 
was used. In this model coherence effects were implicitly included in 
the formalism of the parton cascade.
The Lund string fragmentation \cite{LUND}, 
as implemented in JETSET 7.4 \cite{JET}, was used for the
hadronisation phase.

The GRV-HO \cite{GRV} parameterisation 
was used for the MEPS data set. 
For the CDMBGF event sample the 
MRSD$-^{\prime}$ \cite{MRSD} parton density parameterisation for the
proton was used. These parameterisations describe 
the HERA measurements of the proton structure function 
F$_2$ \cite{GH,F294,Z93F2} reasonably well. 

Monte Carlo samples of
DIS events containing $D^\ast$ mesons (all decay modes) 
corresponding to 5.5\,pb$^{-1}$ were generated.
The CDMBGF sample was used for cross section determinations
and the MEPS sample for systematic studies.
The efficiency for selecting DIS events where a $D^\ast$ has been produced
is determined as a function of $Q^2$ and $y$.  
For the kinematic region selected, 
an average event selection efficiency of 75\% is found, increasing
from about 60\,\% to 90\,\% as $Q^2$ increases from 5 to 100 GeV$^2$
and from about 65\,\% to 85\,\% as $y$ increases from 0 to 0.7.

\section{\bf {\boldmath $D^\ast$} Reconstruction}

The tracks of charged particles 
are reconstructed using the CTD. 
The single hit efficiency of the chamber is greater than 95\%. 
The efficiency for assigning hits to tracks depends on several
factors, for example the \ptr\,of the track and the number of 
nearby charged particles. In addition the 45$^\circ$ inclination
of the drift cells, which compensates for the Lorentz angle, 
introduces some asymmetry in the chamber response for positive
and negative particles, particularly at low \ptr. 
The reconstructed tracks used are 
required to have more than 20 hits, 
a transverse momentum \ptr$\,>\,$0.125\,GeV and a polar angle
between $20^\circ < \theta < 160^\circ$.
In terms of pseudorapidity, $\eta = -\log (\tan (\theta / 2))$, this angle 
corresponds to $\left| \eta \right| < 1.75 $.
This is the region where the tracking detector
response and systematics are sufficiently understood.
For those tracks with \ptr$\,>$\,0.125\,GeV and 
$\left|\,\eta\,\right|<1.75$ the track reconstruction efficiency 
is greater than 94\%.

\subsection{\bf {\boldmath $D^\ast$}  Identification}
$D^\ast$ production is investigated in the decay channel (1). The 
tight kinematic constraint on the 
$D^{\ast +}\rightarrow D^0 \pi^+_s$ decay limits the
momentum of the decay products to 
just 40\,MeV
in the $D^\ast$ rest frame. This fact allows one to measure 
the mass difference $\, M(D^\ast) \, - \, M(D^0)$ more accurately 
than the measurement of the $D^\ast$\,mass itself. In practice it leads
to a prominent signal in the 
$\Delta M \, = \, M(K\pi\pi_s) \, - \, M(K\pi)$
distribution, in an 
otherwise highly suppressed region of phase space.

 The $D^\ast$ reconstruction procedure consists of two steps. 
 First a $D^0$ pre-candidate is formed by taking all combinations of
 pairs of oppositely charged tracks and assuming 
 each track in turn to correspond to a kaon or a pion.
 If the $K\pi $ invariant mass of the track combination lies 
 between 1.4 and 2.5\,GeV, the track pair is considered
 to be a $D^0$ pre-candidate. 
 In the second step this 
 $D^0$ pre-candidate is combined with a third track,
 which has the sign of charge opposite to that of the kaon candidate.
 The third track is assumed to be a pion 
(the so called soft pion, $\pi_s$). 
If the mass difference, $\Delta M$,
is below 180\,MeV, the three tracks form a $D^\ast$ pre-candidate.

To reduce the combinatorial background and restrict the analysis to a kinematic
region where the detector reconstruction efficiency
is acceptable, the following requirements are applied:

\begin{itemize}

\item 
The spatial resolution of the CTD does not allow the $D^0$ decay vertex to be 
distinguished from the primary vertex.
Therefore, we require all tracks to 
be associated with the reconstructed primary vertex of the event.
\item The kaon and pion candidates from the $D^0$ decay
must have transverse momenta
greater than 0.4\,GeV. This reduces the combinatorial background in the $D^0$ 
reconstruction step. 
\item $D^\ast$ pre-candidates must have transverse momenta 
  in the range 1.3~$<$~\ptr~$<$~9.0\,GeV
  and directions of flight away from
  the beams, $\left|\eta\right|~<~1.5$.
  The lower \ptr\ limit is due to the very small acceptance 
  for $D^\ast$ with \ptr\ below 1.3\,GeV which results from 
  the \ptr($\pi_s$)$>$0.125\,GeV cut. 
  The higher \ptr\ limit is due to the lack of statistics.
  Note that these cuts limit the actual acceptance to $y>~0.015$.
\end{itemize}
The analysis is restricted to those combinations 
(called $D^\ast$ candidates)
which pass the above requirements, have 
a $K\pi $ invariant mass in the range 
1.8\,$<\,M(K\pi)~<$~1.92\,GeV and have a mass difference in the range 
143\,$<\, \Delta M\,<$\,148\,MeV. 
Fig.\,\ref{fig:fig_bgf+}b shows the distribution
of all DIS events which have
a $D^\ast$ candidate in the ($x,Q^2$) plane. 

\begin{figure}
\epsfig{file=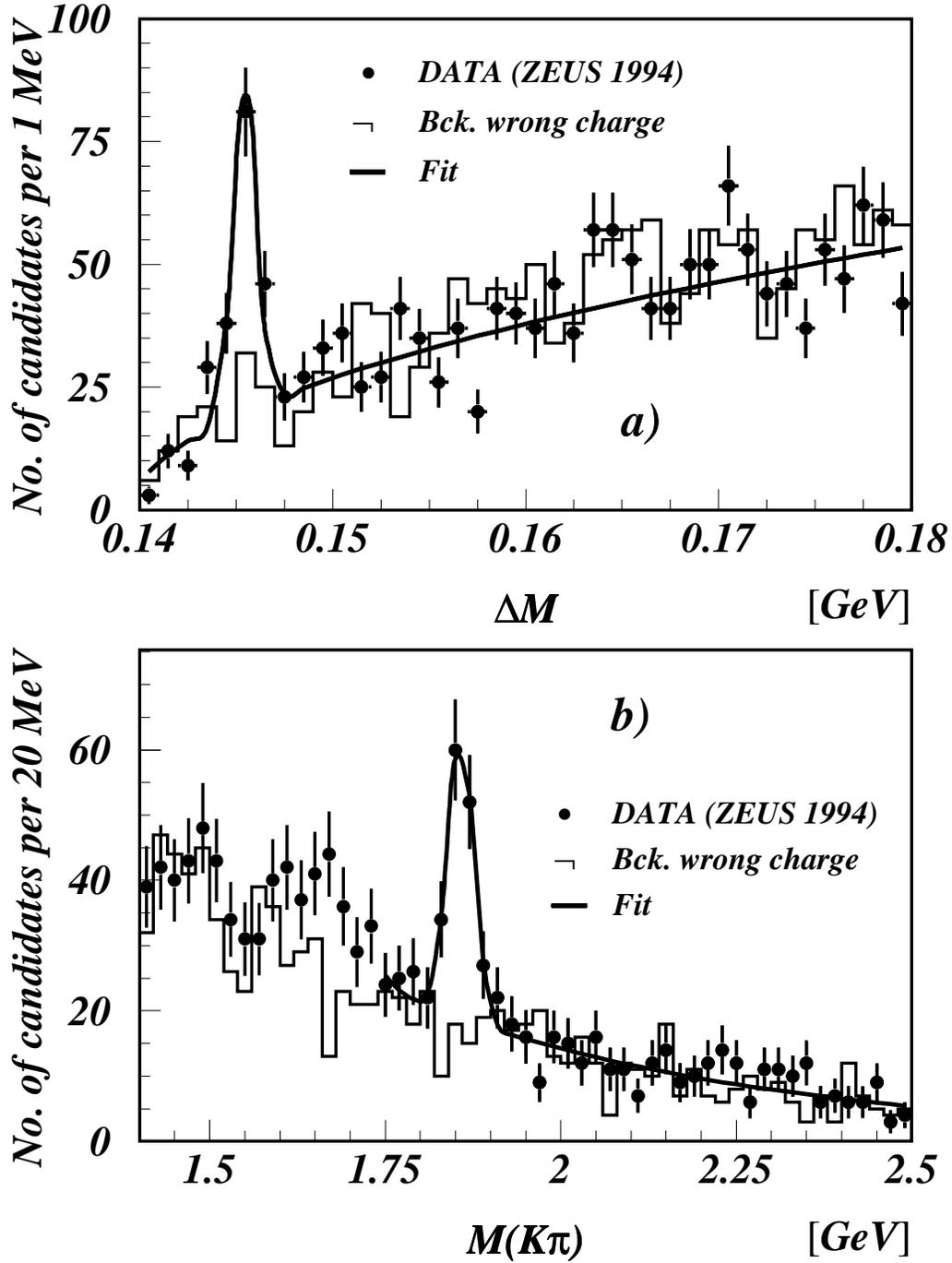,%
bbllx=2cm,bblly=150pt,bburx=525pt,bbury=600pt,height=16cm}
\caption{\label{fig:mass} \em 
(a) $\Delta\,M$ mass distribution for
$K\pi$ combinations in the $M(K\pi)$ signal region 
(1.80~$<~M(K\pi)~<$~1.92\,\GeV), full dots, and for the wrong charge combinations, 
solid histogram.
(b) $M(K\pi)$ mass distribution for 
the $K\pi\pi_s$ combinations in the $\Delta\,M\,=\,M(K\pi\pi_s)\,-\,M(K\pi)$ 
signal region (143~$<~\Delta~M~<$~148\,MeV), full dots, and
for the wrong charge combinations, solid histogram.
The solid lines in both 
figures show the result from the fits (see text for details).
}
\end{figure}

Figs.\,\ref{fig:mass}a,b show the 
resulting $\Delta M$ ($M(K\pi)$)
spectrum for those pre-candidates with  $M(K\pi)$ ($\Delta M$)
inside the corresponding signal region. Clear signals above the
combinatorial background are seen around 
the nominal $M(D^\ast)\,-\,M(D^0)$\, and $M(D^0)$ values. 
The $\Delta M$ spectrum is fitted with the maximum likelihood method 
by a Gaussian shaped signal plus a background of the 
form $dN/d\Delta\,M=a\,(\Delta\,M-m_{\pi})^b$ in the 
140-180\,MeV mass region where $a$ and $b$ are free parameters.
This fit yields $M(D^\ast)\,-\,M(D^0)$\,=\,145.44\,$\pms$\,0.09\,MeV, 
in good agreement with
the PDG \cite{PDG} value of 145.42\,$\pms$\,0.05\,MeV. 
The width of the signal is 0.65 $\pms$ 0.10 MeV.
The excess of events seen in Fig.~\ref{fig:mass}b
in the range of masses between 1.5 and 1.7 GeV is mainly due to the
decay $ D^0 \rightarrow K^- \pi^+ \pi^0$, where the $\pi^0$
is not measured. 
A fit of the $M(K\pi)$ spectrum in the
mass range 1.7\,$<~M(K\pi)~<$\,2.5\,GeV to an exponential curve
plus a Gaussian gives a value for the $D^0$ mass
of 1858~\pms~3\,MeV, slightly below 
the PDG\cite{PDG} value of 1864.6~$\pms$~0.5\,MeV. 
The width of the signal is 19~\pms~3 MeV.
The fit result is shown by the solid curve in
Fig.~\ref{fig:mass}b.
The signal region in $\Delta M$
extends from 143 to 148\,MeV. 
The background under the signal
is estimated in two independent ways: 
\begin{itemize}
\item 
counting the number of combinations which use 
pairs of tracks with the same charge 
for the $D^0$ pre-candidate and fulfill the above 
requirements (wrong charge combinations, 
shown in Fig.~\ref{fig:mass}a as the solid histogram);
\item using 
a control region  2.0\,$<M(K\pi)<$\,2.5\,GeV 
instead of the signal region 
1.80 \,$<M(K\pi)<$\,1.92\,GeV
where the events from the control 
region have been normalized to
the number of events with 155\,$< \, \Delta M\,<$\,180\,MeV.
\end{itemize}
The number of background candidates is taken as the weighted mean 
of the two estimates.
The number of reconstructed $D^\ast$ candidates is taken 
to be the number of entries in the signal region minus 
the number of background candidates. 
For $5<Q^2<100$\,GeV$^2$, 
$y<0.7$ in the restricted kinematic region 
$1.3<\ptr(\ds)<9.0$\,GeV 
and  $|\eta(\ds)|<1.5$ the result is $122\,\pms\,17$ 
$D^\ast$ mesons above a background of $95\,\pms\,8$.
The $\Delta M$ and\,$D^0$ mean value and
width of the signals predicted by the NC DIS MC sample agree well with the 
data. The same is true for the background shape
and the signal to background ratio.
For the measurement of the $\ptr(\ds) ,\,\eta(\ds) ,\,Q^2\,$and$\,W$ 
dependence, the data sample in each variable has been divided
in three bins. The number of $D^\ast$ obtained in each 
bin is determined following the same procedure described above.

\subsection{\bf {\boldmath $D^\ast$} reconstruction efficiency}
   The $D^\ast$ reconstruction efficiency is determined
using MC simulations. 
DIS events 
with $c\bar{c}$ production by photon-gluon fusion
were generated using two different MC 
models: AROMA\,2.1\,\cite{AROMA} and HERWIG\,5.8\,\cite{HERWIG}.
AROMA is a MC model for heavy 
flavour production. It is based on the following 
ingredients: $(i)$ the complete matrix elements in LO
for the PGF process $\gamma^\ast\,g\,
\rightarrow\,c\,\bar{c}$
(taking into account the mass of the charm quark and the full electroweak
structure of the interaction), $(ii)$ gluon emission from the $c \bar{c}$ 
system in a parton shower approach, $(iii)$ initial state parton showers, 
$(iv)$ hadronisation with the Lund string model \cite{LUND} as 
implemented in JETSET\,7.4\,\cite{JET} and heavy flavour decay. 
In addition bremsstrahlung emission of gluons (some of 
which may split perturbatively into two gluons or $q\bar{q}$ pairs) is
also included.

HERWIG is a general purpose QCD MC event generator for 
high energy hadronic processes.
Here it is used for simulating $\gamma^{\ast}\,g\,
\rightarrow\,c\,\bar{c}$ in LO. In addition, 
leading-logarithm parton showers were included in the simulation. 
Fragmentation into hadrons is modeled with a cluster fragmentation 
model which takes into account the charm quark mass.

The parameters of the MC programs were set to their default values.
In particular, the charm quark mass ($m_c$) was set to 
$1.35$\,GeV for AROMA and $1.8$\,GeV for HERWIG. 
For the parton densities of the proton the parameterisations
of MRSD$-^\prime$ \cite{MRSD} and MRSA \cite{MRSA} were considered.

Those events containing
at least one charged $D^\ast$, decaying into $D^0\,\pi^+$ with subsequent decay
$D^0 \rightarrow K^-\pi^+$\,, were processed through the standard ZEUS detector
and trigger simulation programs and through the event reconstruction package.
Approximately 5000 MC events from the 
samples generated and processed in this way, passed all the 
selection criteria.
Both MC reproduced equally well the shapes of the uncorrected
data distributions relevant for this study.
These events were used to determine the
$D^\ast$ reconstruction efficiency as a function of \ptr(\ds)\,and\, $\eta(\ds)$.  
For each $D^\ast$ candidate, the efficiency in a given (\ptr(\ds), $\eta(\ds)$) bin
is the ratio of the number of reconstructed $D^\ast$ 
in the bin to the number of generated $D^\ast$ in the bin.    
Both AROMA and HERWIG MC samples give compatible results and
were combined for the final efficiency determination.
The $\eta(\ds)$ resolution is less than 0.01 in units of pseudorapidity.
The \ptr(\ds)\ resolution is about 2\% of the bin width 
chosen for the figures; none of these show systematic shifts. 
The $D^\ast$ reconstruction efficiency in the kinematic range
considered varies between 20\% for low \ptr(\ds)\ and 70\% for \ptr(\ds)\,
$\ge$\,5.8\,GeV. The efficiency varies in $ \eta $  from
30\% near $\left|\,\eta\,\right| = 1.5$ 
to 50\% for $D^\ast$ moving transversely
to the beam direction ($\eta = 0$). 
The average $D^\ast$ reconstruction efficiency is about 38\%.
As described in section 4.3, the event selection efficiency is determined
using the HERACLES MC. The convolution of the event selection efficiency
and the $D^\ast$ reconstruction efficiency gives an overall detection 
efficiency of approximately 30\% for $D^\ast$ decaying via (1).

\section{\bf Results}
\subsection{\bf {\boldmath $D^\ast$} fractional momentum distribution }

To investigate the charm production mechanism, the 
   distribution of the fractional momentum of the $D^\ast$ in 
   the $\gamma^\ast p$ system 
   ($x_{D^\ast}= {2 |\vec{p}\,\,^\ast_{D^\ast}| \over W }$) is studied.
   The PGF process produces 
   a $c \bar{c}$ pair that, in the $\gamma^\ast p$ system, recoils against the 
   proton remnant. In contrast, in the flavor-excitation 
   process, a single $c$ quark from 
   the proton sea is scattered off the proton, flying, in the $\gamma^\ast p$
   system, in a direction opposite to the proton remnant. Since, in general, the  
   $D^\ast$ carries a large fraction  of the momentum of the 
   parent charm quark 
   \cite{PDG}, clear differences are expected between the $x_{D^\ast}$ 
   distributions from the two production mechanisms. The distribution is expected to be
   centered at $x_{D^\ast} < 0.5$
   for PGF and peaked at high $x_{D^\ast}$ values for direct production.  
   Figure~\ref{fig:fig_bgf+}c
   shows the normalized $x_{D^\ast}$ differential distribution measured in the 
   data and the AROMA prediction (based on PGF). For comparison we also show 
   the prediction by LEPTO6.1  with only direct production from charm.
   Both MC use the proton density of MRSD$-^\prime$\,\cite{MRSD}. 
   The shape of our data distribution is compatible with PGF, in accord
   with the H1 result\,\cite{MEMEO}.

\subsection{\bf Cross sections in a restricted {\boldmath \ptr(\ds)} and {\boldmath $\eta(\ds)$} kinematic region}

Differential cross sections  as well as cross sections integrated over
the kinematic region $1.3<\ptr(\ds)<9.0$\,GeV and $\left|\eta(\ds)\right|<1.5$
are presented in this section. The $e^+p \rightarrow e^+ \dspm X$ differential
cross sections are corrected for the efficiencies
of the selection criteria
as well as for the branching ratios
$B(D^{\ast +}\rightarrow D^0 \pi^+_s)\,
\times\,B(D^0 \rightarrow K^- \pi^+)\,=\,0.0262\,\pm\,0.0010$ \cite{PDG}. 

\begin{figure}
\epsfig{file=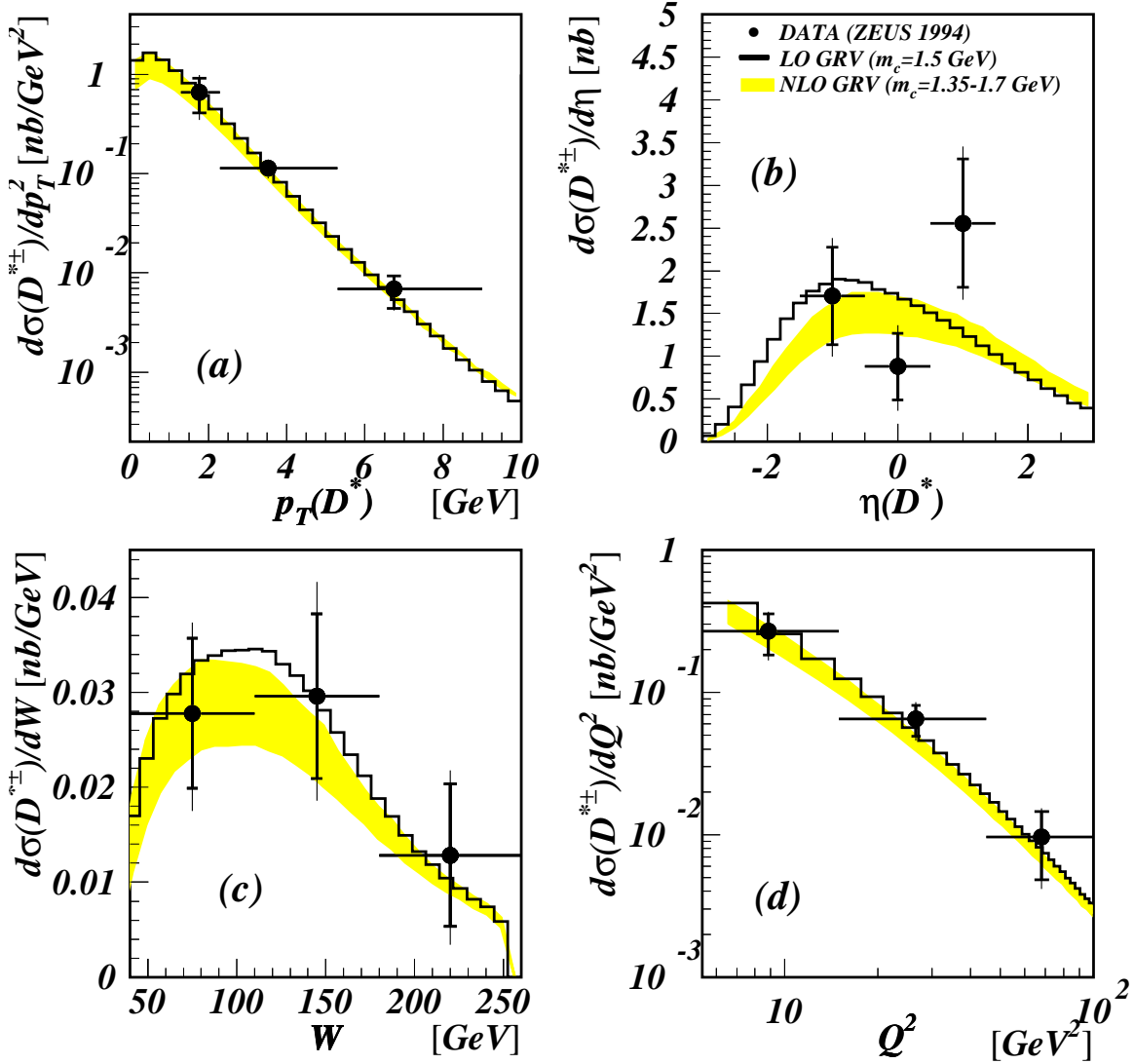,%
bbllx=0cm,bblly=150pt,bburx=530pt,bbury=625pt,height=14cm}
\caption{\label{fig:ds_dares} \em 
Differential $e^+p \rightarrow e^+\dspm X$ cross sections 
for $5<Q^2<100$\GeV$^2$, $y<0.7$ in the restricted kinematic region 
$1.3<\ptr(\ds)<9.0$\,GeV and $|\eta(\ds)|<1.5$ as a function of 
\ptr(\ds) (a), $\eta(\ds)$ (b), $W$ (c) and $Q^2$ (d).
The inner error bars show the statistical errors and the outer ones
correspond to the statistical and systematic errors added in quadrature.
The horizontal bars represent the bin widths. 
The NLO QCD prediction for different charm quark masses 
is shown by the band (see text).
The LO prediction for the GRV(LO) gluon density
is shown by the histogram (see text).
The predicted cross sections in (a) and (b) are 
shown without the \ptr(\ds)\,and $\eta(\ds)$ cuts respectively. }
\end{figure}

Fig.~\ref{fig:ds_dares} shows 
the differential 
$e^+p \rightarrow e^+ \dspm X$ cross section
as a function of a) \ptr(\ds), b) $\eta(\ds)$, c) $W$ and d) $Q^2$ in the kinematic
region defined above. 
The inner error bars are the statistical errors and the outer ones 
show the statistical and systematic errors added in quadrature.
The data points in the \ptr(\ds) and $Q^2$ distributions are shown
at the positions of the average values of an exponential and
polynomial fit, respectively, for a given bin.

The \dspm\, differential cross section $d\sigma/dp{_T}^2(\ds)$
exhibits an exponential falloff in $p{_T}(\ds)$ and $d\sigma/d\eta(\ds)$ 
is approximately flat. The shape of the $W$ dependence of 
the cross section predicted for
photon-gluon fusion at low $W$ is determined by the \ptr(\ds)\,$>$\,1.3\,GeV and
the $ \left|\eta(\ds)\right|\,<\,1.5$ cuts. The falloff at large $W$ is mainly due to
the virtual photon flux.
The \dspm\, cross section drops steeply as $Q^2$ increases.

The systematic uncertainties arise from several effects.
The main uncertainties coming from DIS event selection and
$D^\ast$ reconstruction are:
the transverse momentum cut of the soft pion (7\%), 
the range used for the $K\,\pi$ invariant mass (5\%) and 
the positron energy cut (5\%).
The differences in acceptances evaluated from different MC
samples correspond to an uncertainty
in the integrated $e^+p \rightarrow e^+ \dspm X$ cross section
at the 7\% level. The QED radiative effects contribute 
3\,$\%$ to the systematic error.
The systematic error attributed to the branching ratios is 4\%.
The normalisation uncertainty due to the determination of
the luminosity and trigger efficiency is 2\%.
No significant
contribution to the systematic error from the primary vertex 
requirement was found.
Adding the contributions in quadrature, a total 
systematic error of 15\% is obtained 
in the determination of the integrated cross section. 
For the differential distributions the systematic errors 
are estimated bin by bin.

The results of the NLO analytic calculation 
of \cite{HARRIS}
using the  GRV\cite{GRV94} NLO gluon density, $xg(x,Q^2)$, of the proton
is shown as a band in Fig.~\ref{fig:ds_dares}, where
the upper (lower) limit corresponds to a charm quark mass of 1.35\,(1.7)\,GeV.
The calculation is based exclusively on the PGF process and no charm produced 
from the proton sea is considered. 
The GRV NLO parton density gives a reasonable description of the proton structure 
function $F_2$ measured at HERA ~\cite{GH,F294} and treats the charm quarks
as massive particles, consistent with the ansatz of \cite{HARRIS}.
Charm quarks were hadronized according to the Peterson fragmentation 
function\,\cite{PETER}
with $\epsilon_c\,=\,0.035\,\pms\,0.009$ \cite{AEC}
and the probability for a charm quark to fragment into $D^\ast$ ,
$P(c\rightarrow D^\ast)$, was assumed to be 0.26\,\pms\,0.02 \cite{AEC}.
Both the renormalisation and the factorisation
scales were chosen to be $\mu\,=\,\sqrt{Q^2\,+\,4\,m_c^2}$.
In this calculation the largest uncertainty 
is related to that of the charm quark mass.
The predicted differential cross section for \dspm\, production
as a function of a)~\ptr(\ds)~and b)~$\eta(\ds)$ 
is shown in Fig.~\ref{fig:ds_dares} outside the 
restricted kinematic region selected for the data.
The NLO analytic calculation  
reproduces well the shapes of the \ptr(\ds),\,$W$ and $Q^2$ distributions
and is consistent
with the $\eta(\ds)$ dependence in the restricted kinematic region.

Below, unless explicitly stated otherwise, the LO and NLO analytic calculations
are those from \cite{HARRIS} and use respectively the LO and NLO
gluon density parameterisations of GRV \cite{GRV94} and
a charm quark mass $m_c = 1.5$ GeV. 

In Fig.~\ref{fig:ds_dares} the data are also compared with 
the LO analytic calculation, shown as histograms.
The LO calculation performed with the MC programs used
in the acceptance calculation and described in section 5.2 agrees
well with the LO analytic calculation.
Note that the MC calculation includes parton showers in addition to the LO matrix
elements.
The LO MC and LO analytic calculations 
describe the shapes of the \ptr(\ds)\,, $Q^2$ and
$W$ dependences equally well.

Integrated over the region
$5<Q^2<100$\,GeV$^2$ and $y~<~0.7$,
the $e^+p \rightarrow e^+ \dspm X$ cross section
for \dspm\, in the restricted kinematic range $1.3<\ptr(\ds)~<9.0$\,GeV and 
$\left|\eta(\ds)\right|<1.5$ is 
$$ \sigma_{p_T,\eta}\,(e^+p\,\rightarrow\,e^+\,\dspm\,X)\,=\,
5.3\,\pms\,1.0\,\pms\,0.8\,{\rm nb}$$
where the first error is statistical and the second error is systematic.
The $e^+p \rightarrow e^+ \dspm X$ cross section
represents about 4~$\%$ of the DIS cross section 
(calculated from structure functions determined by fits
to global data \cite{MRSD}) in the same
$Q^2$ and $y$ ranges.
Table\,\ref{tab:dstar_results} compares the measured cross section
with the LO and NLO predictions.
The integrated cross sections predicted in NLO 
is in reasonable agreement with the data.
The LO MC and analytic calculation agree with the data,
although their sensitivity to the parameters involved
in the program ($m_c$, $xg(x,Q^2)$, $\epsilon_c$ and $\mu$) is higher 
than in the case of NLO calculations.

\subsection{\bf Integrated charm cross section and {\boldmath $F_2^{c\bar{c}}$} }

 The DIS inclusive cross section for charm production, 
$e^+p\rightarrow e^+\,c\bar{c}\,X$, $\sigma^{c\bar{c}}$,
can be expressed in terms of
$F_2^{c\bar{c}}$ using:
\begin{equation}
{{d^2 \sigma^{c\bar{c}}} \over {dx dQ^2}} = 
{{2\pi\alpha^2}\,\over\,{Q^4x}} [(1+(1-y)^2)\,F_2^{c\bar{c}}(x,Q^2)
 \,-\,y^2\,F_L^{c\bar{c}}(x,Q^2)]
\end{equation}
In the standard model, the contribution from $Z$-boson exchange 
is expected to be small in the ($Q^2,x$) range of the present analysis
and therefore the $F_3^{c\bar{c}}$ contribution
is neglected. The $F_L^{c\bar{c}}$ 
contribution has been estimated \cite{RIEM} to be smaller 
than 2\% and, therefore, no correction has been applied.

In order to estimate $\sigma^{c\bar{c}}$ and to evaluate
$F_2^{c\bar{c}}$ as a function of $Q^2$ and $y$, the measurements
must be extrapolated to the full \ptr(\ds)\, and $\eta(\ds)$ range. 
$\sigma^{c\bar{c}}$ is obtained from the integrated $D^\ast$ cross section 
using:
\begin{equation}
\sigma(e^+p\rightarrow e^+\,c\bar{c}\,X)= {1 \over 2} \cdot 
{{\sigma(e^+p\rightarrow e^+\,D^\ast \,X)} \over {P(c\rightarrow D^\ast)}}.
\end{equation}
According to the extrapolation with the NLO calculation,
about 50\,\% (at $Q^2=45$\,GeV$^2$ ) to 65\,\% (at $Q^2=7$\,GeV$^2$ )
of the $D^\ast$ production 
is found to be outside the restricted kinematic region. 
Using the LO analytic calculation for the extrapolation results
in similar cross sections.

For the determination of the integrated cross section the data 
are divided into two $Q^2$ intervals, namely 
5$\,<\,Q^2\,<\,10\,$GeV$^2$ and 10$\,<\,Q^2\,<\,100\,$GeV$^2$. 
The extrapolated acceptance is calculated using the NLO prediction of
GRV\cite{GRV94} with $m_c$=1.5\,GeV,\,$\mu=\sqrt{Q^2\,+\,4m_c^2} $, 
$P(c \rightarrow D^{\ast+})$\,=\,0.26\,\pms\,0.02 
and ${\epsilon}_c\,=\,0.035\,\pms\,0.009$ \cite{AEC}. 
The integrated charm cross section, 
$\sigma\,(e^+p\,\rightarrow\,e^+\,c\bar{c}\,X)$,
for $y\,<\,0.7$, is 
$13.5\,\pms\,5.2\,\pms\,1.8\,^{+1.6}_{-1.2}\,$nb
for 5$<Q^2<$10\,GeV$^2$,
and $12.5\,\pms\,3.1\,\pms\,1.8\,^{+1.5}_{-1.1}$nb
for 10$\,<\,Q^2\,<\,100\,$GeV$^2$. 
The first error gives the statistical, the second 
the experimental systematic uncertainty and the third 
the model dependent uncertainty. The latter is
studied by varying the relevant parameters of the model
used for extrapolation, namely  $m_c$ (from 1.35 to 1.7\,GeV),
$\mu$ (from 2$m_c$ to $2\,\sqrt{Q^2\,+\,m_c^2}$), $\epsilon_c$ (from
0.025 to 0.045), and $xg(x,Q^2)$ 
(GRV \cite{GRV94},\,MRSG \cite{MRSG}, 
CTEQ3 \cite{CTEQ}, MRSA$^\prime$\cite{MRSA}) 
\footnote{Note that while GRV (and also CTEQ4F3 and MRRS, see below)
follow the charm quark treatment of
\cite{HARRIS} based on PGF, the CTEQ3 and MRS distributions 
treat charmed quarks as massless quarks with a parametrized density
contributing above a certain threshold in $Q^2$ and
therefore are not completely consistent with the program of\,\cite{HARRIS}.} 
and also includes the error in the probability for $c \rightarrow D^\ast~ $.
An uncertainty between $-$12\,\% and +9\,\% 
is obtained for both $Q^2$ ranges.

The integrated charm cross sections as well as the corresponding predictions 
from the LO and NLO analytic calculations are also listed 
in Table\,\ref{tab:dstar_results}.
The NLO predictions are about one standard deviation smaller than the measured value
in the first $Q^2$ interval (from 5 to 10\,GeV$^2$) but in agreement 
in the second $Q^2$ interval (from 10 to 100\,GeV$^2$).
The charm cross section obtained for 10$\,<\,Q^2\,<\,100\,$GeV$^2$
is in reasonable agreement with the recent H1 measurement in the same
$Q^2$ region.

\begin{table}[htbp]
\begin{center}             
\begin{tabular}{|c|c|c|c|}   
\hline
\hline
 & $\sigma_{p_T,\eta}$ [nb] &  $\sigma_1$ [nb] &  $\sigma_2$ [nb]\\ 
\hline
\hline
  \multicolumn{1}{|c|}{Data} & 5.3\,\pms\,1.0\,\pms\,0.8 &$13.5\,\pms\,5.2\,\pms\,1.8\,^{+1.6}_{-1.2}\,$   & $12.5\,\pms\,3.1\,\pms\,1.8\,^{+1.5}_{-1.1}$\\ 
\hline
\ \ \ AROMA & 4.57  & 12.6  & 14.2  \\
\hline
\ \ \ LO Analytic calculation & 4.79 & 11.0 & 12.4  \\
\hline
\ \ \ NLO Analytic calculation & 4.15 & 9.4 & 11.1  \\
\hline
\hline
\end{tabular}
\caption{ 
\label{tab:dstar_results}
\it
{\small
The measured $e^+p \rightarrow e^+\dspm X$ cross 
section for $5<Q^2<100$\,GeV$^2$, 
$y<0.7$ in the restricted kinematic region 
$1.3<\ptr(\ds)<9.0$\,GeV 
and $|\eta(\ds)|<1.5$ ($\sigma_{p_T,\eta}$);
the extrapolated $e^+p \rightarrow e^+c\bar{c} X$ cross section
for $y<0.7$ and for $5<Q^2<10$\,GeV$^2$ ($\sigma_1$)
and $10<Q^2<100$\,GeV$^2$ ($\sigma_2$).
The measured cross sections are compared with the NLO analytic
calculation for the GRV NLO gluon density 
as well as with the LO MC and LO analytic calculations.
The statistical uncertainties of the NLO analytic
calculation have about 0.01 the size of the uncertainties of the data points.}
}
\end{center}
\end{table}

\begin{figure}
\epsfig{file=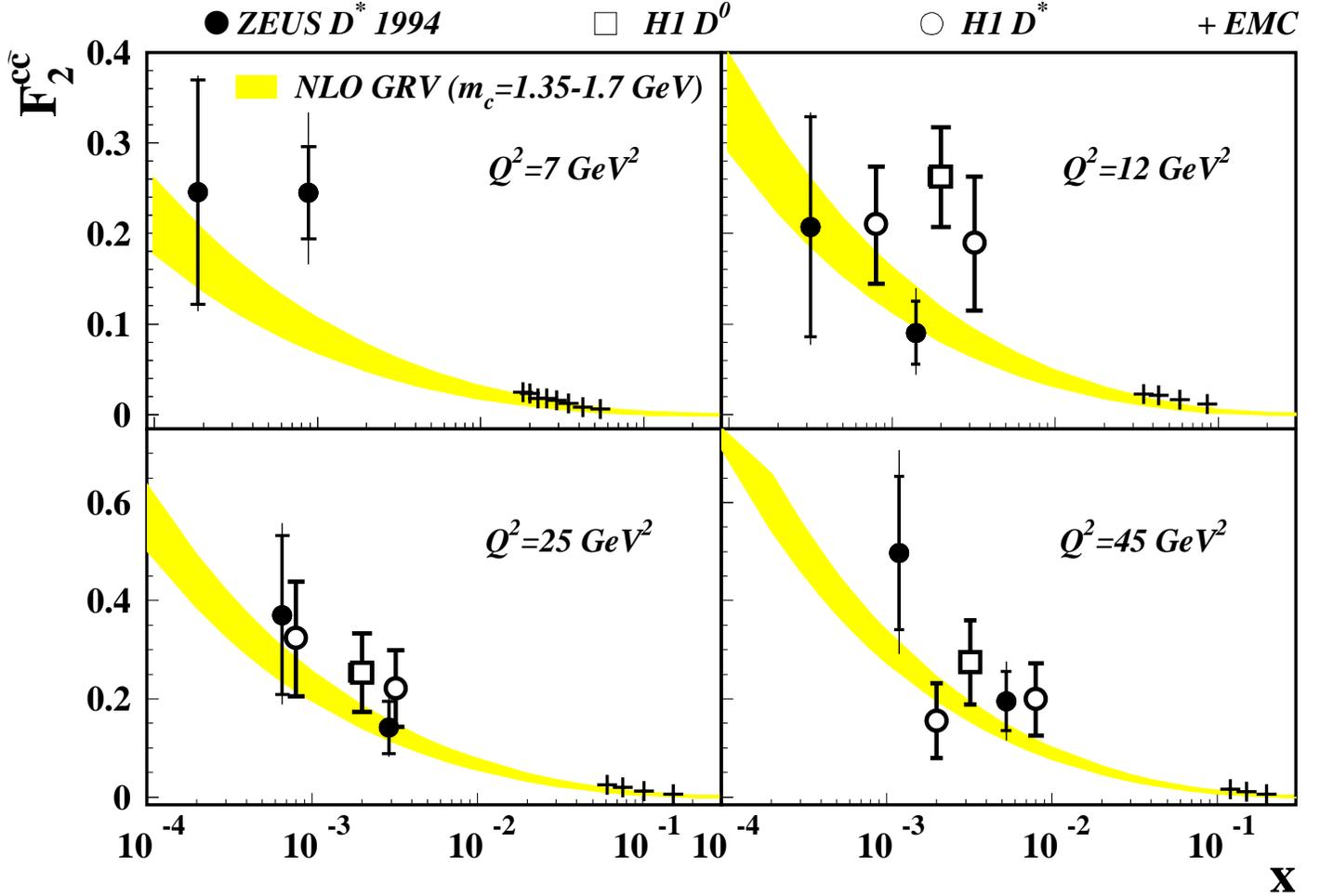,%
bbllx=2cm,bblly=150pt,bburx=525pt,bbury=600pt,height=16cm}
\caption{\label{fig:f2_c} \em 
The charm contribution, $F_2^{c\bar{c}}$, to the proton structure
function $F_2$ as derived from the inclusive $D^\ast$(ZEUS and H1)
and $D^0$(H1) production 
compared with the NLO QCD predictions based on the
GRV parton distribution
using different charm quark masses for $Q^2$=7, 12, 25 and 45 \GeV$^2$
The upper (lower) limit of the band corresponds to a charm quark mass of 1.35\,(1.7)\,\GeV (see text).
The results from the EMC collaboration are shown as crosses.
For the ZEUS data, the inner error bars show the statistical 
errors and the outer ones
correspond to the statistical and systematic errors added in quadrature.
The error bars from H1  
show the statistical and systematic errors added in quadrature.
The error bars for EMC are within the symbol.
 }
\end{figure}

Fig.~\ref{fig:f2_c} shows 
the resulting $F_2^{c\bar{c}}$ values as a function of $x$
for the different $Q^2$ and $y$ bins as defined in Fig.~\ref{fig:fig_bgf+}b. 
The H1 measurements \cite{MEMEO}
are also shown; they were taken in the same $Q^2$ bins. The two
sets of data are in good agreement. 
Also included in Fig.~\ref{fig:f2_c} are the data from 
the EMC fixed target experiment\,\cite{EMC} which were measured
at $x$ values between 0.02 and 0.3.
The charm contribution, $F_2^{c\bar{c}}$, to the proton
structure function is seen to rise
by about one order of magnitude from the
high $x$ region covered by the fixed target experiments to the 
low $x$ region measured by H1 and ZEUS.
The NLO analytic calculation 
for $F_2^{c\bar{c}}$  is shown in
Fig.~\ref{fig:f2_c} as a band
(the upper (lower) limit corresponds to a charm quark mass of 1.35\,(1.7)\,GeV).
In the region of our measurements, 
the use of CTEQ4F3\,\cite{CTEQ4} or MRRS\,\cite{MRRS} gluon densities give
results which are within 5\,\% of those using GRV. 
In the present $Q^2$ range, $F_2^{c\bar{c}}$ scales roughly with the
input gluon density in NLO perturbative QCD.   
The measured rise of $F_2^{c\bar{c}}$ from the high $x$
to the low $x$ region is reasonably described by NLO perturbative QCD.

Comparing with our measurement of $F_2(x,Q^2)$ \cite{F294},
we observe that the ratio $F_2^{c\bar{c}}/F_2$ is about 25\% for the entire
($Q^2,x$) range of the present analysis.

\section{\bf Summary and conclusions}

We have measured the $D^\ast$ differential and integrated 
$e^+p \rightarrow e^+\dspm X$ cross sections
for deep inelastic scattering at $\sqrt{s}=300$\,GeV
with 5\,$<Q^2<$100\,GeV$^2$ and $y~<~0.7$
in the restricted kinematic 
region of $1.3<\ptr(D^\ast)<$\,9.0\,GeV and $\left|\eta(D^\ast)\right|<1.5$.
The integrated \dspm\, cross section is  
measured to be 5.3\,\pms\,1.0\,(stat.)\,\pms\,0.8\,(syst.)\,nb.

The shape of the $D^\ast$ fractional momentum distribution 
in the $\gamma^\ast p$ rest system, $x_{D^\ast}$, shows
that the PGF mechanism prediction agrees well
with the data for DIS charm
production in this kinematic range. 

A QCD analytic calculation in NLO with the NLO GRV
gluon density 
reproduces the shapes of the \ptr(\ds)\, $W$ and $Q^2$ 
distributions and is consistent
with the $\eta(\ds)$ dependence in the restricted kinematic region.
The predicted cross sections are in reasonable agreement with the data.
We have used QCD calculations to extrapolate the \dspm\, 
cross section measured in the restricted \ptr(\ds),\,$\eta(\ds)$ region 
to the full
region and estimated the integrated charm cross section and the charm
contribution $F_2^{c\bar{c}}$ to the proton structure function $F_2$.
When compared
to the fixed target measurements (performed at large $x$) $F_2^{c\bar{c}}$
is found to rise as $x$ decreases. 
The rise is described by NLO QCD calculations when using
a gluon density consistent with that extracted from the 
scaling violations in the 
proton structure function $F_2$ measured at HERA.
Such a gluon density distribution is also compatible with our previous
measurements of $D^\ast$ in photoproduction.

\section*{Acknowledgements}

The strong support and encouragement by the DESY Directorate have been
invaluable.
The experiment was made possible by the inventiveness 
and diligent efforts of the HERA machine group.

The design, construction and installation of 
the ZEUS detector have been made possible 
by the ingenuity and dedicated efforts of 
many people from the home institutes who are not listed here. Their 
contributions are acknowledged with great appreciation.
We acknowledge the support of the DESY computing and 
network services.
We would like to thank J. Smith for valuable discussions and 
S. Riemersma for helpful comments and for 
providing us with the NLO analytic $F_2^{c\bar{c}}$ calculation.

We warmly acknowledge B. Harris for the close collaboration 
during the last stage of the analysis 
and for providing us with
the LO and NLO analytic charm cross section calculations.


\end{document}